\documentclass[preprint,showpacs,showkewords,preprintnumber,amsmath,amssymb]{revtex4}
 \usepackage{tipa}
 \usepackage{amsmath}
 \usepackage{txfonts}
 \usepackage{amssymb}
 \usepackage{graphicx,epsfig}
 \usepackage{bm}
 \begin{document}

 \title{ Propose economical and stable lepton mass matrices with texture zeros }
 \author{Shu-jun Rong}\email{rongshj@snut.edu.cn}

 \affiliation{Department of Physics, Shaanxi University of Technology, Hanzhong, Shaanxi 723000, China}

 \begin{abstract}
There are many viable combinations of texture zeros in lepton mass matrices. We propose an economical and stable mass texture. Analytical and numerical results on mixing parameters and the effective mass of neutrinos are obtained. These results satisfy new constraints from neutrinos oscillation experiments and cosmological observations. Our proposition reveals that in the complex forest of neutrinos mixing models, a simple and robust one is still possible.
\end{abstract}


 \maketitle
\section{introduction}
\label{intro}
Neutrinos oscillation is one of most mysterious phenomena in particle physics, which reveals that neutrinos are massive and mixing structure of them are nontrivial. However, how to interpret the origin of masses and special mixing pattern of neutrinos is still an open question. As useful phenomenological scenarios, special textures of lepton mass matrices and corresponding flavor groups are introduced to predict mixing parameters of neutrinos. In particular, lepton mass matrices with texture zeros \cite{1,2,3,4,5,6,7,8} or structures alike \cite{9,10,11,12} are widely employed in mixing models of neutrinos, which can be realized in Abelian or non-Abelian flavor groups, see Refs \cite{13,14,15} for example. As an important progress, the recent Ref\cite{16} has performed a complete survey of possible combinations of texture zeros in mass matrices of charged leptons and Majorana or Dirac neutrinos. There are so many types of possible mass matrices with texture zeros that we don't know what is the true texture chosen by nature. In order to evaluate importance of classes of texture zeros, we propose two criteria: First, the mass matrices with texture zeros should be as economical as possible; Second, predictions of texture-zeros models should be stable under perturbations of mass matrices. We consider that an economical model of texture zeros can give clear dependences of observables on parameters of lepton mass matrices. And texture-zeros models with robust predictions are more sustainable, because they can react to the progress of neutrinos experiments by introduction small modifications or perturbations.

In this letter, we propose a special combination of texture zeros of leptons mass matrices with an economical and stable structure. The mass matrix of charged leptons is Hermitian with Fritzsch texture \cite{17,18,19}, i.e.,
\begin{equation}
\label{eq:1}
 M_{l}=
 \left(
   \begin{array}{ccc}
     0 & A_{l} & 0 \\
    A^{\ast}_{l} & 0 & B_{l} \\
     0 & B^{\ast}_{l} & C_{l} \\
   \end{array}
 \right),
 \end{equation}
 where $C_{l}$ is real and $A_{l}$, $B_{l}$ are complex parameters.
And the mass matrix of Majorana neutrinos with one zero, i.e.,
\begin{equation}
\label{eq:2}
 M_{\nu}=
 \left(
   \begin{array}{ccc}
    a+c & b & 0 \\
    b & a & c \\
     0 & c & a+b \\
   \end{array}
 \right),
 \end{equation}
 where $a$ is complex and $b$, $c$ are real parameters. As for $M_{l}$, because of its special structure, resorting to a diagonal phase matrix, it can be transformed to a real symmetric matrix with three parameters determined by masses of charged leptons. So it can be diagonalized by a matrix with a simple analytical expression. As for $M_{\nu}$, it is magic, which predicts a trimaximal mixing pattern (TM$_{2}$)~\cite{20} if the mass matrix of charged leptons is diagonal. The structure of $M_{\nu}$ is characterised by the ratio $b/c$. So it can be diagonalized by a matrix with one real parameter. Furthermore, if $a$ is a real parameter, $M_{\nu}$ would be more economical. However, in this case, ranges of mixing parameters would be smaller. So we introduce $a$ as a complex parameter in this paper.

  In following sections, the analytical expression of the lepton mixing matrix is given. Numerical predictions of the mixing parameters are obtained by scanning parameter space of mass matrices. Stabilities of these predictions are examined by introduction of random perturbations that break texture zeros of lepton mass matrices. Finally, a summary is presented.

\section{Analytical framework}
In this section, we consider the analytical expression of mixing matrix and dependence of observables on parameters in lepton mass matrices.
\subsection{Analytical expression of mixing matrix}
The lepton mixing matrix with the standard parametrization is expressed as
\begin{equation}
\label{eq:3}
U=diag\left(
        \begin{array}{ccc}
          e^{i\delta_{e}}, & e^{i\delta_{\mu}}, & e^{i\delta_{\tau}} \\
        \end{array}
      \right)\cdot V\cdot diag\left(
        \begin{array}{ccc}
          e^{-i\varphi_{1}/2}, & e^{-i\varphi_{2}/2}, & 1 \\
        \end{array}
      \right)
\end{equation}
where $\delta_{e,~\mu,~\tau}$ are unphysical phases, $\varphi_{1,~2}$ are Majorana phases, $V$ is written as~\cite{21}
\begin{equation}
\label{eq:4}
V=
\left(
\begin{array}{ccc}
 c_{12}c_{13} & s_{12}c_{13} & s_{13}e^{-i\delta} \\
 -s_{12}c_{23}-c_{12}s_{13}s_{23}e^{i\delta} & c_{12}c_{23}-s_{12}s_{13}s_{23}e^{i\delta} & c_{13}s_{23} \\
s_{12}s_{23}-c_{12}s_{13}c_{23}e^{i\delta} & -c_{12}s_{23}-s_{12}s_{13}c_{23}e^{i\delta} & c_{13}c_{23}
\end{array}
\right),
\end{equation}
where $s_{ij}\equiv\sin{\theta_{ij}}$, $c_{ij}\equiv\cos{\theta_{ij}}$, $\delta$ is the Dirac CP-violating phase.
This standard mixing matrix can be obtained through $U=U^{+}_{l}U_{\nu}$, where $U_{l}$ in our model comes from diagonalization of the Hermitian mass matrix $M_{l}$, i.e.
\begin{equation}
\label{eq:5}
U^{+}_{l}M_{l}U_{l}= diag\left(
        \begin{array}{ccc}
          m_{e}, & m_{\mu}, & m_{\tau} \\
        \end{array}
      \right),
\end{equation}
$U_{\nu}$ comes from diagonalization of mass matrix of Majorana neutrinos, i.e.
\begin{equation}
\label{eq:6}
U^{T}_{\nu}M_{\nu}U_{\nu}= diag\left(
        \begin{array}{ccc}
          m_{1}, & m_{2}, & m_{3} \\
        \end{array}
      \right).
\end{equation}
In detail, $U_{l}=P_{l}O_{l}$, where
\begin{equation}
\label{eq:7}
P_{l}=diag\left(
        \begin{array}{ccc}
          e^{i\alpha}, & e^{i\beta}, & 1 \\
        \end{array}
      \right),
\end{equation}
where $\alpha=arg[A_{l}]+arg[B_{l}]$, $\beta=arg[B_{l}]$,
elements of $O_{l}$ are written as~\cite{22,23}
 \begin{equation}
 (O_{l})_{11}=[\frac{x_{l}-z_{l}}{(1+x_{l})(1-z_{l})(x_{l}-z_{l}+x_{l}z_{l}))}]^{1/2},\nonumber
 \end{equation}

  \begin{equation}
 (O_{l})_{12}=-i[\frac{(1+z_{l})x^{3}_{l}}{(1+x_{l})(x_{l}+z_{l})(x_{l}-z_{l}+x_{l}z_{l}))}]^{1/2},\nonumber
 \end{equation}

 \begin{equation}
 (O_{l})_{13}=[\frac{(1-x_{l})z^{3}_{l}}{(1-z_{l})(x_{l}+z_{l})(x_{l}-z_{l}+x_{l}z_{l}))}]^{1/2},\nonumber
 \end{equation}

 \begin{equation}
 (O_{l})_{21}=[\frac{(x_{l}-z_{l})}{(1+x_{l})(1-z_{l})}]^{1/2},\nonumber
 \end{equation}

  \begin{equation}
 (O_{l})_{22}=i[\frac{x_{l}(1+z_{l})}{(1+x_{l})(x_{l}+z_{l})}]^{1/2},\nonumber
 \end{equation}

  \begin{equation}
 (O_{l})_{23}=[\frac{z_{l}(1-x_{l})}{(1-z_{l})(x_{l}+z_{l})}]^{1/2},\nonumber
 \end{equation}

  \begin{equation}
 (O_{l})_{31}=-[\frac{x_{l}z_{l}(1-x_{l})(1+z_{l})}{(1+x_{l})(1-z_{l})(x_{l}-z_{l}+x_{l}z_{l}))}]^{1/2},\nonumber
 \end{equation}

  \begin{equation}
 (O_{l})_{32}=-i[\frac{z_{l}(1-x_{l})(x_{l}-z_{l})}{(1+x_{l})(x_{l}+z_{l})(x_{l}-z_{l}+x_{l}z_{l}))}]^{1/2},\nonumber
 \end{equation}

\begin{equation}
 (O_{l})_{33}=[\frac{x_{l}(1+z_{l})(x_{l}-z_{l})}{(1-z_{l})(x_{l}+z_{l})(x_{l}-z_{l}+x_{l}z_{l}))}]^{1/2},
 \end{equation}
where $x_{l}=m_{e}/m_{\mu}$, $z_{l}=m_{e}/m_{\tau}$. Because mass hierarchy of charged leptons is strong, approximation of $O_{l}$ is
\begin{equation}
\label{eq:9}
O_{l}\approx\left(
        \begin{array}{ccc}
          1 & -ix^{1/2}_{l} & z^{3/2}_{l}x^{-1}_{l} \\
         x^{1/2}_{l} & i & z^{1/2}_{l}x^{-1/2}_{l} \\
          -z^{1/2}_{l} & -iz^{1/2}_{l}x^{-1/2}_{l} & 1 \\
        \end{array}
      \right).
\end{equation}
Furthermore, mass parameters of $M_{l}$ could also be expressed with masses of charged leptons, i. e.,~\cite{22}
 \begin{equation}
 \label{eq:10}
|A_{l}|=(\frac{m_{e}m_{\mu}m_{\tau}}{m_{e}-m_{\mu}+m_{\tau}})^{1/2},\nonumber
 \end{equation}
 \begin{equation}
|B_{l}|=[\frac{(m_{e}-m_{\mu})(m_{\mu}-m_{\tau})(m_{e}+m_{\tau})}{m_{e}-m_{\mu}+m_{\tau}}]^{1/2},\nonumber
 \end{equation}
 \begin{equation}
C_{l}=m_{e}-m_{\mu}+m_{\tau}.
 \end{equation}

As for neutrinos, $U_{\nu}=O_{\nu}P_{\nu}$, where
\begin{equation}
O_{\nu}=\left(
          \begin{array}{ccc}
            O_{1\nu} & O_{2\nu} & O_{3\nu}\\
          \end{array}
        \right),
\end{equation}
\begin{equation}
 O_{1\nu}=N_{1\nu}\left(
   \begin{array}{c}
      -1+x_{\nu}-\sqrt{1-x_{\nu}+x^{2}_{\nu}} \\
      x_{\nu}+\sqrt{1-x_{\nu}+x^{2}_{\nu}} \\
     1\\
   \end{array}
 \right),
 ~O_{2\nu}=\frac{1}{\sqrt{3}}\left(
   \begin{array}{c}
      1 \\
      1 \\
     1\\
   \end{array}
 \right),~O_{3\nu}= N_{3\nu}\left(
   \begin{array}{c}
      -1+x_{\nu}+\sqrt{1-x_{\nu}+x^{2}_{\nu}} \\
     x_{\nu}-\sqrt{1-x_{\nu}+x^{2}_{\nu}} \\
     1\\
   \end{array}
 \right),
\end{equation}
where $x_{\nu}=b/c$, $N_{1\nu}$ and $N_{3\nu}$ are normalization factors expressed as
 \begin{equation}
 \begin{array}{c}
  N_{1\nu}=[\sqrt{1+(-1+x_{\nu}-\sqrt{1-x_{\nu}+x^{2}_{\nu}})^{2}+(x_{\nu}+\sqrt{1-x_{\nu}+x^{2}_{\nu}})^{2}} ]^{-1}, \\
  N_{3\nu}=[\sqrt{1+(-1+x_{\nu}+\sqrt{1-x_{\nu}+x^{2}_{\nu}})^{2}+(x_{\nu}-\sqrt{1-x_{\nu}+x^{2}_{\nu}})^{2}} ]^{-1}.
\end{array}
\end{equation}
And the diagonal phase matrix $P_{\nu}$ is expressed as
\begin{equation}
P_{\nu}=diag\left(
              \begin{array}{ccc}
                e^{-i\phi_{1}/2}, & e^{-i\phi_{2}/2}, & e^{-i\phi_{3}/2} \\
              \end{array}
            \right),
\end{equation}
where $\phi_{1}=arg[a-\sqrt{b^2-bc+c^2}]$, $\phi_{2}=arg[a+b+c]$, $\phi_{3}=arg[a+\sqrt{b^2-bc+c^2}]$.
Employing expressions of $O_{l}$, $P_{l}$, $O_{\nu}$, and $P_{\nu}$ listed above, elements of the lepton mixing matrix can be written as
\begin{equation}
\label{eq:15}
U_{ij}=(O^{+}_{l})_{i1}(O_{\nu})_{1j}e^{-i(\phi_{j}/2+\alpha)}+(O^{+}_{l})_{i2}(O_{\nu})_{2j}e^{-i(\phi_{j}/2+\beta)}+(O^{+}_{l})_{i3}(O_{\nu})_{3j}e^{-i\phi_{j}/2}.
\end{equation}
Obviously, $\phi_{j}s$ are just relevant to Majorana phases of the mixing matrix.

\subsection{Dependence of observables on parameters in lepton mass matrices}
The mixing angles of leptons can be obtained from comparison of expression of $U_{ij}$ in Eq.~\ref{eq:15} with the form of $U$ in the standard parametrization in Eq.~\ref{eq:3} and Eq.~\ref{eq:4}. In detail, for $\theta_{13}$,
\begin{equation}
\label{eq:16}
\sin\theta_{13}=|U_{13}|=|(O^{+}_{l})_{11}(O_{\nu})_{13}e^{-i\alpha}+(O^{+}_{l})_{12}(O_{\nu})_{23}e^{-i\beta}+(O^{+}_{l})_{13}(O_{\nu})_{33}|.
\end{equation}
Because $O_{l}$ is dependent on masses of charged leptons that are determined by experiments very well, $\theta_{13}$ in fact is dependent on three real effective mass parameter: $\alpha$, $\beta$, and $x_{\nu}$. The same observation can also hold for $\theta_{23}$, $\theta_{13}$ and $\delta$. Therefore, three mixing angles and Dirac CP violating phase are not independent to each.

As for masses of neutrinos, i.e.,
\begin{equation}
\label{eq:17}
  m_{1}=|a-\sqrt{b^2-bc+c^2}|, ~~
  m_{2}= |a+b+c|,~~
  m_{3}=|a+\sqrt{b^2-bc+c^2}|,
\end{equation}
they are dependent on four real parameters of $M_{\nu}$. And the effective mass of neutrinoless double-beta decay, i.e.,
\begin{equation}
\label{eq:18}
m_{ee}=m_{1}U^{2}_{11}+m_{2}U^{2}_{12}+m_{3}U^{2}_{13},
\end{equation}
is dependent on all 6 real parameters in $M_{l}$ and $M_{\nu}$.

\section{Predictions of lepton mass matrices and their stabilities}
\subsection{Predictions of lepton mass matrices}
We take masses of charged leptons from Particle Data Group~\cite{24}:
\begin{equation}
  m_{e}=0.51 ~MeV, ~~
 m_{\mu}=105.658 ~MeV,~~
  m_{\tau}=1776.86 ~MeV.
\end{equation}
Then using the Mixing-Parameters-Tools package~\cite{25,26} , we scan 6 real parameters in lepton mass matrices, i.e., arg[$A_{l}$], arg[$B_{l}$], arg$[a]$, $|a|$, $b$, $c$. The numerical results are shown in Fig. 1, including predictions for mixing angles ($\theta_{13}$, $\theta_{12}$, $\theta_{23}$), Dirac CP violating phase $\delta$, mass ratios ($m_{1}/m_{2}$, $m_{2}/m_{3}$), the mass of the lightest neutrino $m_{1}$, sum of masses of neutrinos $\sum m_{i}$, the effective mass of neutrinoless double-beta decay $m_{ee}$. Some comments are given as follows:\\
$\bullet$ ~~ Different from the case where both of mass matrices of charged leptons and neutrinos are of the form of Fritzsch texture~\cite{27}, the octant of $\theta_{23}$ is uncertain in our case.\\
$\bullet$ ~~ Because of the strong mass hierarchy of charged leptons, the range of the Dirac CP violating phase $\delta$ is narrow, i.e. , $0^{\circ}\pm 50^{\circ}$ or $180^{\circ}\pm 50^{\circ}$. \\
$\bullet$ ~~ The the upper limit of the mass of the lightest neutrino is small, i.e., about 23.5 meV. So is that of the effective mass of double-beta decay i.e., about 24 meV. Accordingly, the rate of neutrinoless double-beta decay may be too small to detect in the present stage~\cite{29}.\\
$\bullet$ ~~ The mass ordering of neutrinos is normal. And the hierarchy of them is not strong. so the upper limit of the sum of masses of neutrinos is also small, i.e. approximating 102~meV, which satisfies new stringent constrains from cosmological observations, e.g., $\sum m_{i}<146~meV$ at the 2$\sigma$ level in Ref.~\cite{30}.

\begin{figure}[tbp]
\centering 
\includegraphics[width=.45\textwidth]{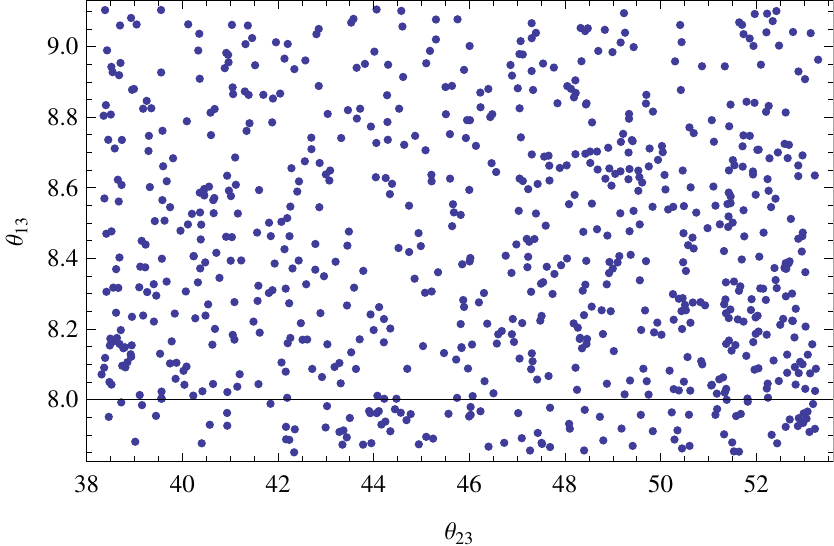}
\hfill
\includegraphics[width=.45\textwidth]{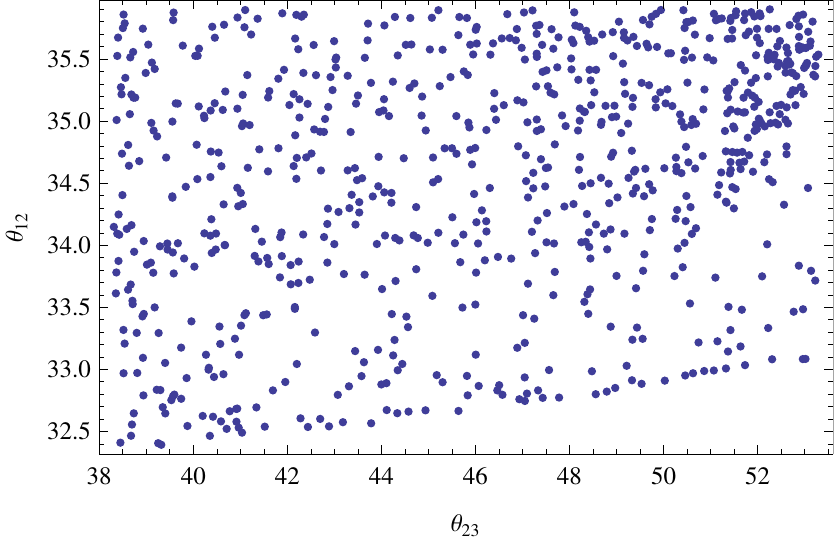}
\hfill
\includegraphics[width=.45\textwidth]{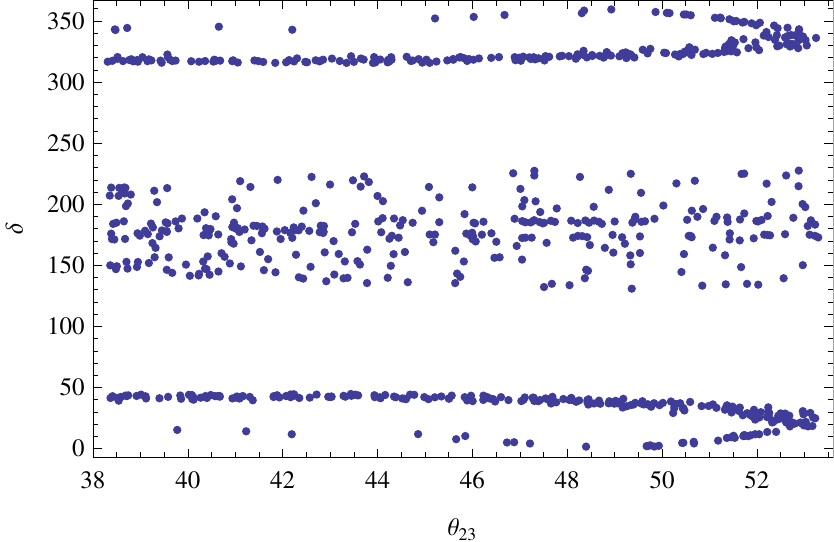}
\hfill
\includegraphics[width=.45\textwidth]{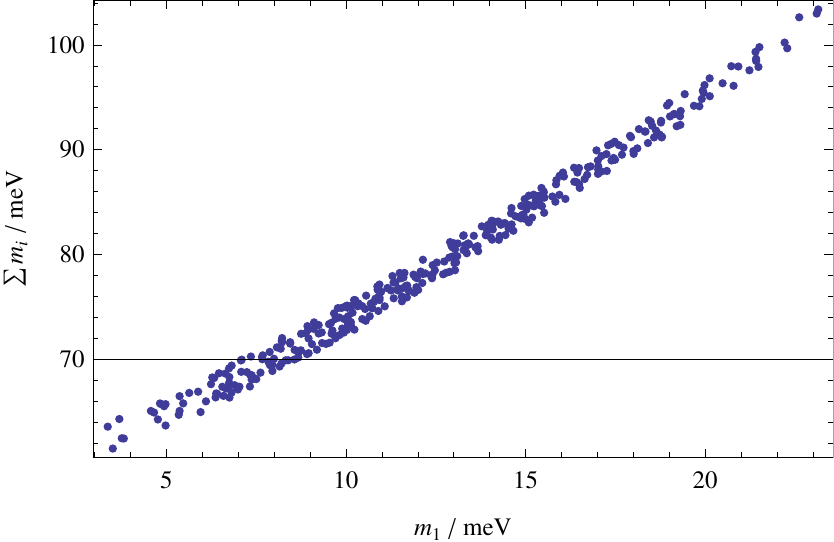}
\hfill
\includegraphics[width=.45\textwidth]{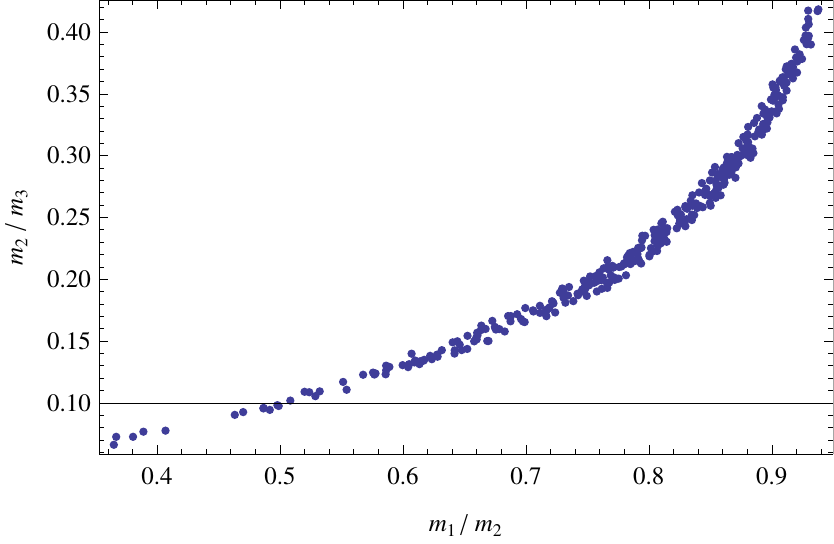}
\hfill
\includegraphics[width=.45\textwidth]{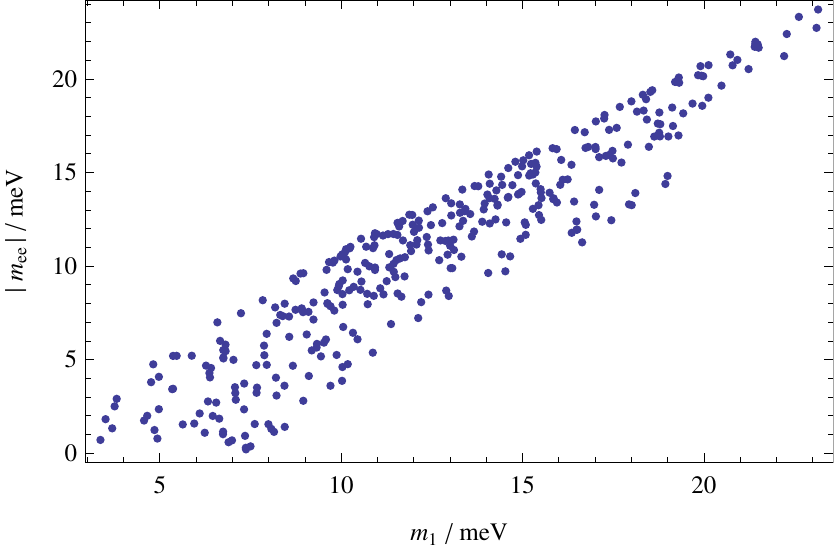}
\hfill
\caption{\label{fig:1} Predictions for mixing angles ($\theta_{13}$, $\theta_{12}$, $\theta_{23}$), Dirac CP violating phase $\delta$, mass ratios ($m_{1}/m_{2}$, $m_{2}/m_{3}$), the mass of the lightest neutrino $m_{1}$, sum of masses of neutrinos $\sum m_{i}$, the effective mass of neutrinoless double-beta decay $m_{ee}$. Here the unit of angles is degree. }
\end{figure}

\subsection{Stabilities of predictions under perturbations breaking zeros}
\begin{figure}[tbp]
\centering 
\includegraphics[width=.45\textwidth]{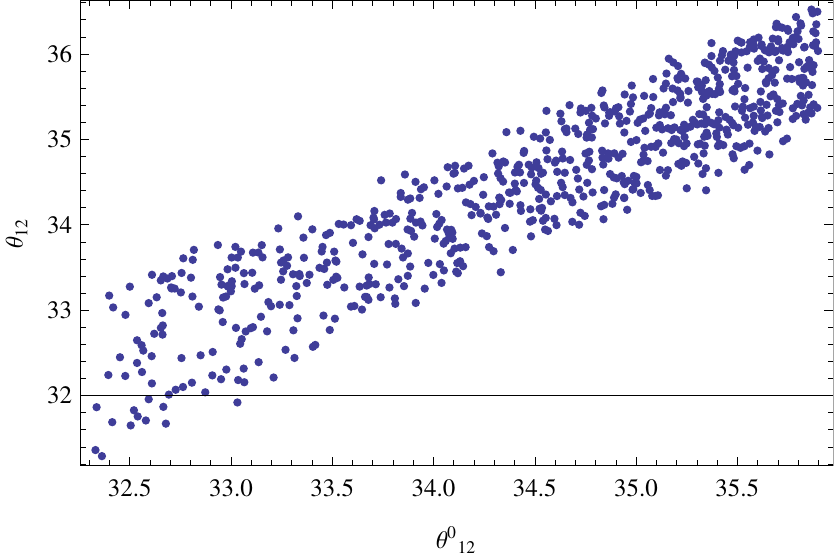}
\hfill
\includegraphics[width=.45\textwidth]{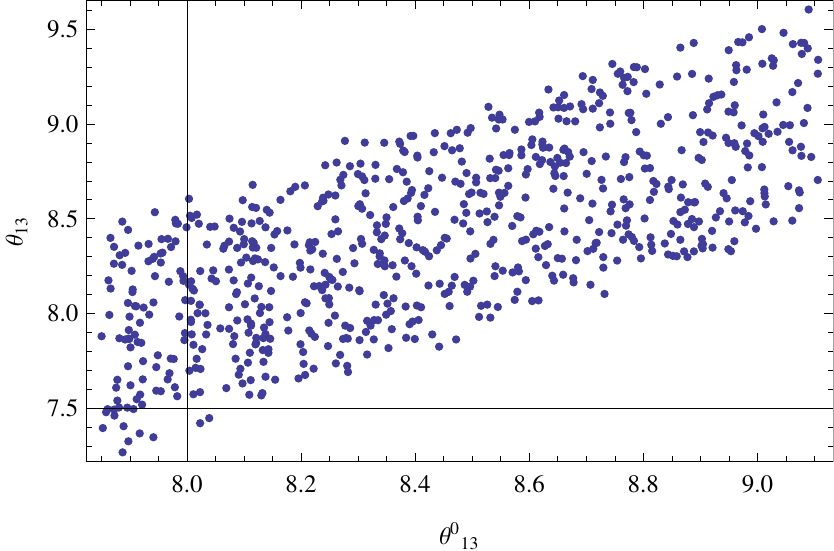}
\hfill
\includegraphics[width=.45\textwidth]{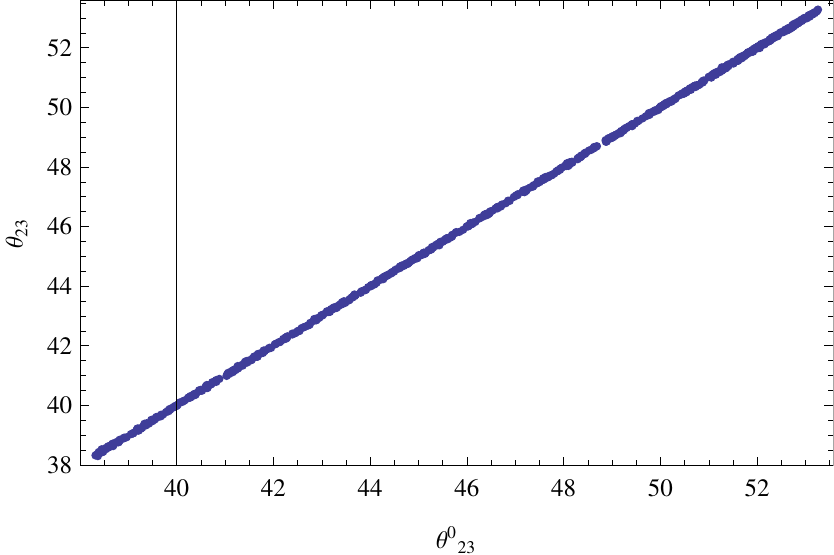}
\hfill
\includegraphics[width=.45\textwidth]{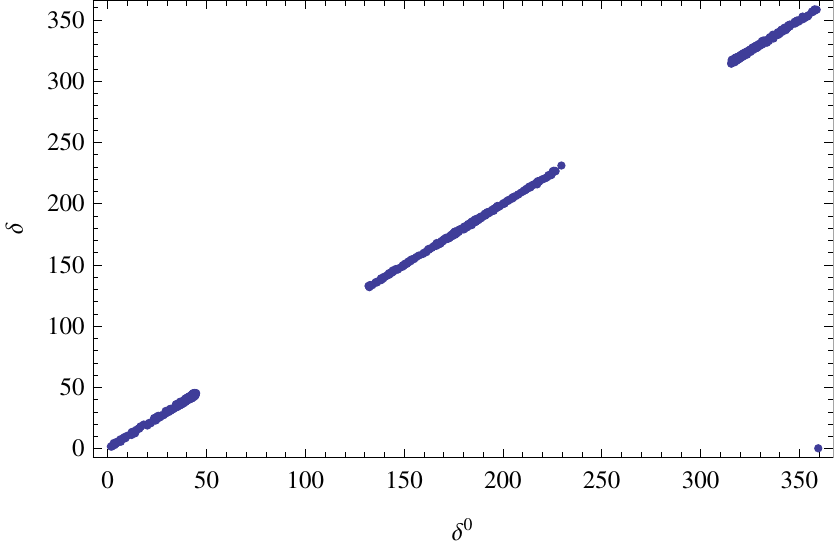}
\hfill
\includegraphics[width=.45\textwidth]{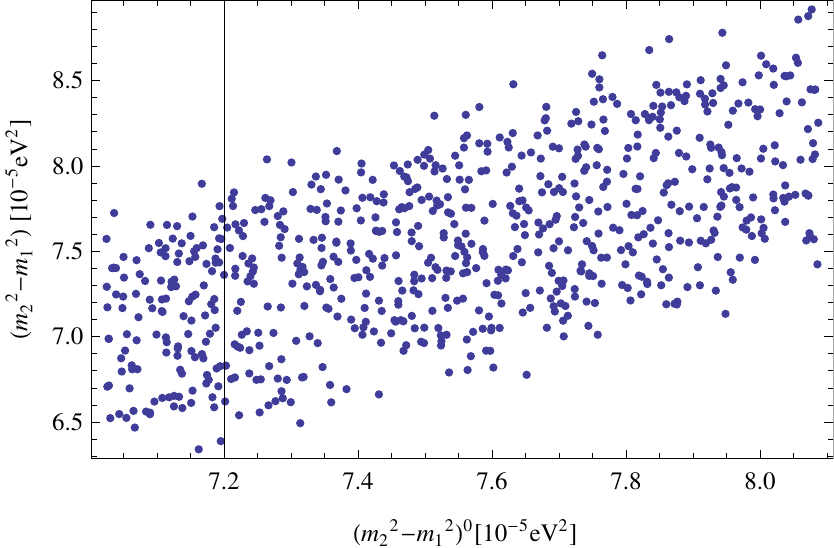}
\hfill
\includegraphics[width=.45\textwidth]{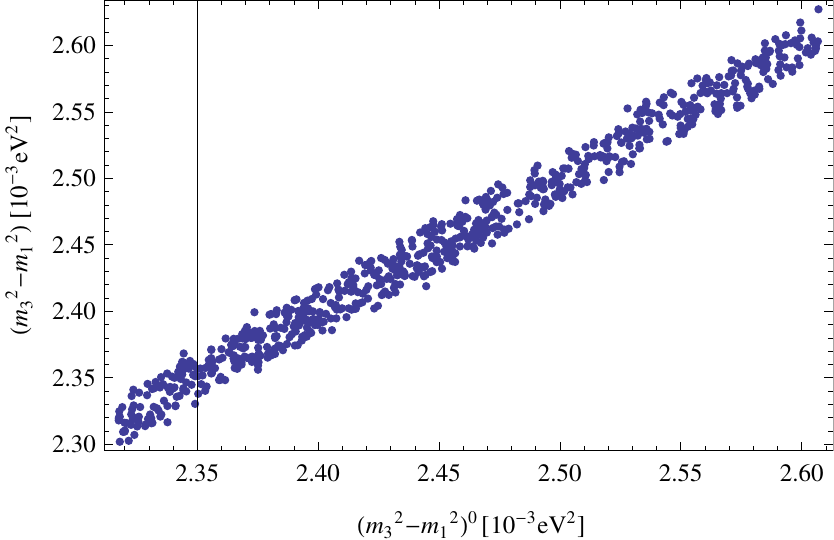}
\hfill
\caption{\label{fig:2} Correlations of neutrinos mixing parameters before and after perturbations. Here the superscripts "0" in the plots denote initial mixing parameters (before perturbations). The initial mixing parameters are in the ranges of 3$\sigma$ from the Ref.~\cite{28} And the unit of angles is degree. }
\end{figure}

The above predictions are obtained on the basis of the special combinations of texture zeros of leptons mass matrices. In order to examine stabilities of these predictions, we introduce perturbations that break zeros in the mass matrices. For charged leptons, the perturbation is proposed as
\begin{equation}
 M_{l}=
 \left(
   \begin{array}{ccc}
     \epsilon_{11} & A_{l} & \epsilon_{13}\\
    A^{\ast}_{l} & \epsilon_{12} & B_{l} \\
     \epsilon^{\ast}_{13} & B^{\ast}_{l} & C_{l} \\
   \end{array}
 \right),
 \end{equation}
where $\epsilon_{11}$ and $\epsilon_{12}$ are real perturbations of order $\mathcal{O}(0.01 m_{e})$, $\epsilon_{13}$ is a complex perturbation of order $\mathcal{O}(0.01m_{\mu})$.
For neutrinos, the mass matrix after the perturbation is
\begin{equation}
 M_{\nu}=
 \left(
   \begin{array}{ccc}
    a+c & b & \epsilon' \\
    b & a & c \\
     \epsilon'  & c & a+b \\
   \end{array}
 \right),
 \end{equation}
where $\epsilon'$ is a real perturbation of order $\mathcal{O}(0.01 |a|)$. Together with initial parameters in lepton mass matrices, these parameters of perturbations are scanned. And plots of correlations of neutrinos oscillation parameters before and after perturbations are shown in Fig.~\ref{fig:2}.
Some comments are given as follows:\\
$\bullet$ ~~ For $\theta_{12}$ and $\theta_{13}$, variations brought by perturbations approximate $1^{\circ}$, while for $\theta_{23}$ and $\delta$ variations are nearly negligible.\\
$\bullet$ ~~ For the mass squared difference $\delta m^{2}_{21}=m^{2}_{2}-m^{2}_{1}$, the relative variation brought by perturbations approximate $10\%$. The relative variation of $\delta m^{2}_{31}=m^{2}_{3}-m^{2}_{1}$ is about approximate $1\%$.\\
$\bullet$ ~~ The mixing parameters after perturbations are still in the 3$\sigma$ ranges of the experiments fit, except marginal data of $\theta_{13}$ and $\delta m^{2}_{21}$. Therefore, the predictions of lepton mass matrices of our special texture are stable under random perturbations that break zeros in them.

\section{Summary}
There are many viable combinations of texture zeros in lepton mass matrices. According to the criteria that the structure of leptons mass matrices should be economical and the prediction of them should be robust, a special combination of texture zeros of mass matrices is proposed. The analytical expression of lepton mixing matrix is obtained, which shows clear dependence of a observable on model parameters. By scanning of parameter-space of mass matrices of leptons, numerical predictions for mixing angles, Dirac CP violating phase, and masses and effective masses of neutrinos are given. These results satisfy constraints from neutrinos oscillation experiments and those from new cosmological observations. These results are also stable under perturbations that break zeros in mass matrices. So, in the complex forest of neutrinos mixing models, a simple and robust one is still possible.

\acknowledgments
The author thank Y.F. Li for helpful discussion. This work is supported by the research foundation of Shaanxi University of Technology under the Grant No. SLGQD-13-10 and the National Natural Science Foundation of China under the Grant No. 11405101.

\end{document}